\documentclass[showpacs,preprintnumbers]{revtex4}
\usepackage{graphicx,amsmath,amssymb}

\newcommand{\half}{\mbox{\small $1 \over 2$}}
\newcommand{\quarter}{\mbox{\small $1 \over 4$}}

\begin{document}

\title{Continuum and lattice heat currents for oscillator chains.}
\author{Onuttom Narayan and A. P. Young}
\affiliation{Department of Physics, University of California,
Santa Cruz, CA 95064}
\date{\today}
\begin{abstract}
We show that two commonly used definitions for the heat current give
different results---through the Kubo formula---for the heat conductivity
of oscillator chains.  The difference exists for finite chains, and is
expected to be important more generally for small structures.  For a
chain of $N$ particles that are tethered at the ends, the ratio of the
heat conductivities calculated with the two currents differs from unity
by $O(1/N).$ For a chain held at constant pressure, the difference from
unity decays more slowly, and is consistent with $O(1/N^\eta)$ with $1 >
\eta > 0.5.$
\end{abstract}
\maketitle
\section{Introduction}
\label{sec:intro}
In linear response theory, transport currents that flow in a system in
response to small gradients in thermodynamic potentials are calculated
in terms of equilibrium autocorrelation functions of these currents
through the Green-Kubo formula~\cite{kubo}. This involves taking the
thermodynamic limit of the autocorrelation functions first, and then
the zero frequency limit; the order in which the limits are taken is
important~\cite{luttinger}. The response function to be calculated
determines the appropriate correlation function: for instance, the
electric and thermal conductivities involve the autocorrelation functions
of the particle and heat currents respectively.

Unlike the particle current, which is defined unambiguously, there
are many possible choices for the energy current, and therefore the
heat current which is a linear combination of the two~\cite{forster}.
Most notably, the heat current can be defined in the continuum or
on a lattice; both are commonly used.  For the lattice current, the
energy of each particle is assigned to the lattice site associated with
the particle. This is commonly used for oscillator chains, and seems
reasonable for crystals (if we neglect lattice defects). The energy
current flowing from the lattice site $i$ to $k$ is
\begin{equation}
j_{ik} = \half {\bf F}_{ik}\cdot ({\bf v}_i + {\bf v}_k)
\end{equation}
where ${\bf F}_{ik}$ is the force exerted on the $k$'th particle
by the $i$'th particle, and ${\bf v}_{i,k}$ are the velocities of
the particles. For the continuum current, commonly used for hard
particle systems or fluids, the energy of each particle resides at its
instantaneous location. As a result, there is an advective part to the
energy current from the motion of particles
\begin{equation}
{\bf j}_{adv}({\bf x}) = \sum_i e_i {\bf v}_i \delta({\bf x} - {\bf x}_i)
\end{equation}
where $e_i$ is the energy of the $i$'th particle. In addition, since the
energy current, $j_{ik}$, between particles $i$ and $k$ flows between
$x_k$ and $x_i$ instead of between their lattice sites, the spatial
integral of $j_{ik}$ (used in the Green-Kubo formula) is also different.

In view of the obvious approximations in the lattice current, is it
merely something that works well for stiff crystals? When computing
the conductivity from equilibrium correlation functions, are the two
currents equivalent in the thermodynamic limit\cite{foot:abh}? This would
not be sufficient for small structures that are specially designed and
cannot be scaled up, for which the conductance has to be used instead
of the conductivity~\cite{fisher,szafer}. Even for large systems, one
has to be careful in proving the equivalence of the two currents if the
transport coefficients are singular, since the thermodynamic limit of
the conductivity does not exist.

Surprisingly, despite the approximations in the lattice current,
one can prove an exact Green-Kubo like formula with it for the
conductivity of a finite chain which has Langevin baths attached to
the terminal particles~\cite{kdn}, using the steady state fluctuation
theorem~\cite{ssft}.  This result has been proved without the fluctuation
theorem and generalized to a number of different implementations of heat
baths~\cite{KDN}. The proof applies to a chain with arbitrary onsite and
interparticle potentials (which may vary down the chain, even though the
proof does not state this). Therefore, it applies to a tethered or free
chain and (by a simple transformation that we shall show in this paper)
a chain at constant pressure.  Is this agreement accidental, and can the
proof of Ref.~\cite{kdn} be extended to the (exact) continuum current
as well?

In this paper, we explain the result by obtaining an interpretation of
the dynamics of the oscillator chain for which the lattice current is
exact; in this interpretation, the Langevin baths are at fixed locations.
We further show through analytical and numerical calculations on such a
chain that the conductivities obtained using the lattice and continuum
currents are not equivalent except in the thermodynamic limit. The
ratio of the two differs from unity by $\sim 1/N$ for a chain with
$N$ particles that is tethered at both ends. For a chain maintained
at constant pressure, the approach to unity is slower. If fit to a
$\sim 1/N^\eta$ form, it is consistent with any $\half < \eta < 1.$ Our
numerics also show that slight changes in the definition of the continuum
heat current, which is a linear combination of the energy and particle
currents, can markedly change the error in the resultant conductivity;
these differences are inconsequential in the thermodynamic limit. The 
conductivities obtained from lattice and continuum currents are 
generally very close for $N\gtrsim 256.$

One can try to explain the error in the Green-Kubo formula using the
continuum current by noting that the size of the system fluctuates in the
continuum interpretation. In the tethered case, the terminal particles
are adjacent to the tethering posts and the percentage fluctuations
in the length are $\sim 1/N,$ while for a chain at constant pressure
they are $\sim 1/\sqrt N.$ This expectation is only partly borne out by
the numerics. Note that, in view of the proof in Refs.~\cite{kdn,KDN}, 
it is the {\it lattice\/} current that gives the heat conductivity accurately.

The rest of this paper is organized as follows. Section II reviews
the derivation of the lattice and continuum heat currents and shows
how the lattice heat current is not an approximation if the oscillator
dynamics are interpreted suitably. Section III calculates the ratio of
the conductivities obtained with the continuum and lattice currents and
how they behave as the thermodynamic limit is approached.  Section IV
presents numerical results with the same for a tethered chain and a
chain at constant pressure.

\section{Transverse and longitudinal interpretations}
\label{sec:transverse}
We consider a one
dimensional chain of particles with nonlinear springs connecting adjacent
particles. (The definitions in this section are standard~\cite{llp}, but we include them for clarity.)
The Hamiltonian is
\begin{equation}
H = \sum_{i=1}^N \half m v_i^2 + \sum_{i=1}^{N-1} U(x_{i+1} - x_i) 
+ \sum_{i=1}^N U_0(x_i)
\label{Ham}
\end{equation}
where $m_i, x_i$ and $v_i$ are the mass, position and velocity of
the $i$'th particle and $U$ is the interparticle potential. $U_0$ is
an onsite potential for each particle. When $U_0 = 0,$ this is a
Fermi-Pasta-Ulam (FPU) chain~\cite{fpu}. The dynamics of this system
are Hamiltonian, with heat baths connected to the first and last
particle through the addition of Langevin damping $-\gamma v_{1, N}$
and noise $\eta_{1,N}(t)$ terms in their equations of motion.  As with
other one-dimensional systems in which interparticle interactions
are momentum conserving~\cite{prosen,ramaswamy}, the FPU chain 
has a singular heat conductivity, with the thermal conductivity
of the chain growing as $\sim N^{1/3}$ as a function of chain
length~\cite{ramaswamy,others,others1,others2}. In higher dimensions or if $U_0\neq 0,$
the singularity in the heat conductivity is eliminated.

If we associate the kinetic energy of each particle with itself, and
the potential energy of each spring as being equally distributed between
its two neighbors, the rate of change of the energy of particle $i$
\begin{equation}
e_i  = \half m v_i^2 +  U_0(x_i) + \half[U(x_{i+1} - x_i) + U(x_{i-1} - x_i)] 
\label{ei}
\end{equation}
is
\begin{equation}
\dot e_i = m_i v_i \dot v_i +U_0^\prime(x_i) v_i + \half F_{i, i-1}(v_{i-1} - v_i)
+ \half F_{i, i+1}(v_{i+1} - v_i)
\end{equation}
where $F_{i,i-1}$ is the force on the $i$'th particle from the $i-1$'th
particle. (There is only one $F$ term in $e_1$ and $e_N$.) Using the
equations of motion, this is equal to
\begin{equation}
\dot e_i = \half F_{i, i-1}(v_{i-1} + v_i) + \half F_{i, i+1} (v_{i+1} + v_i)
+ \delta_{i, 1} (-\gamma v_1 + \eta_1) v_1 
+ \delta_{i, N} (-\gamma v_N + \eta_N) v_N
\label{hdot}
\end{equation}
Therefore the continuum energy current is
\begin{equation}
j^c_E(x, t) = 
\sum_{i=1}^N e_i v_i \delta(x - x_i(t)) + 
\half \sum_{i=2}^N F_{i, i-1}(v_{i-1} + v_i)\theta(x - x_{i-1})\theta(x_i - x)
\label{jcEcomp}
\end{equation}
where the first term comes from the advective transport of the energy
associated with each particle by it. Integrating, the energy current
flowing through the entire system is
\begin{equation}
J^c_E(t) = 
\sum_{i=1}^N e_i  v_i 
+ \half \sum_{i=2}^N F_{i, i-1}(v_{i-1} + v_i)(x_i - x_{i-1}).
\label{Jcont}
\end{equation}
With other choices of how the potential energy of each spring is
distributed, this expression can change by small amounts.
This result can easily be generalized to higher dimensions:
\begin{equation}
{\bf J}^c_E(t) = 
\sum_{i=1}^N e_i  {\bf v}_i 
+ \quarter \sum\sum {\bf F}_{ik}\cdot({\bf v}_k + {\bf v}_i)({\bf x}_i - {\bf x}_k).
\label{Jcont_v}
\end{equation}

On the other hand, if the springs between the particles are very stiff,
each particle only deviates slightly from its lattice position. If 
we neglect the advective term in the heat current and approximate 
$x_i\approx i a$ (where $a$ is the lattice constant), we obtain an 
energy current equal to $F_{i, i-1} (v_{i-1} + v_i)$ between the 
$i-1$'th and $i$'th lattice sites, which integrates to the lattice
energy current
\begin{equation}
J^l_E(t) =\half  \sum_{i=2}^N F_{i, i-1}(v_{i-1} + v_i) a.
\label{Jlat}
\end{equation}
The generalization to higher dimensions is again straightforward. For a
one-dimensional system, in which the ordering of particles is fixed, it is
always possible to view the positions of the particles as displacements
from a reference lattice, but in higher dimensions this is only useful
in a crystalline phase.

By its construction, the lattice energy current $J_E^l$ seems clearly an
approximation; if a Green-Kubo like formula exists for finite oscillator
chains, one would expect it to be only approximate if the lattice current
is used. Despite this, one can prove an exact Green-Kubo type formula
for finite oscillator chains with Langevin baths if the {\it
lattice\/} current is used~\cite{kdn}:
\begin{equation}
\kappa = \frac{1}{k_B T^2 (N-1) a}
\int_0^\infty \langle J_Q(t) J_Q(0)\rangle dt .
\label{kubo}
\end{equation}
In Eq.~(\ref{kubo}) $J_Q(t)$ is the spatially integrated heat current
(defined in the next section) flowing at time $t$ through the chain of $N$
particles at a temperature $T.$ 

As mentioned in the previous section, one can construct an interpretation
of the oscillator chain dynamics for which the lattice current is
exact. Let $x_i = i a + \zeta_i,$ where $\zeta_i$ is the displacement
of the $i$'th particle from its lattice position. We imagine that the
displacement is transverse to the chain instead of along it. If the
potential between neighboring particles is expressed as $U(\zeta_{i+1}
+ a - \zeta_i),$ the dynamics are the same as if the displacements
$\zeta_i$ had been in the longitudinal direction. The energy
current flowing into the Langevin baths is also unchanged.  In this
interpretation, the spatially integrated energy current is $J^l_E$
with no approximations~\cite{foot:hydro}: there is no advective term
in the current, and the energy current between neighboring particles
flows over a distance $a$ instead of $x_i - x_{i-1} = a + \zeta_i -
\zeta_{i-1}.$ (The transverse interpretation is in fact more natural for
the often-studied `phantom' oscillator chains, where particles can pass
through each other but only neighboring lattice sites interact.)

Since the lattice and continuum currents are exact for the transverse and 
longitudinal interpretations respectively, and the heat flowing between
the reservoirs is independent of interpretation, one might hope that 
Eq.(\ref{kubo}) would be valid for both currents. In the rest of this
paper, we shall show that this is not the case except in the thermodynamic
limit.

\section{Lattice and continuum heat currents}
\label{sec:heatcurrents}
The proof in Ref.~\cite{kdn} uses the lattice energy current for the
heat current for a system at zero pressure. Applying a pressure at the
ends of the chain is equivalent to replacing $U(x_i -
x_{i-1})$ with $U(x_i - x_{i-1}) - p (x_i - x_{i-1}).$ From Eq.(\ref{Jlat}),
this changes the heat current to 
\begin{equation}
J_Q^l(t) = \half \sum_{i=2}^N [F_{i, i-1} -p] (v_{i-1} + v_i) a
\label{gauge}
\end{equation}
with which definition the proof of Ref.~\cite{kdn} is extended to systems
at non-zero pressure.
The heat current in the continuum interpretation is defined through Galilean
invariance as~\cite{forster,kubostatmech}
\begin{equation}
J_Q^c(t) = J_E^c(t) - [E + (N-1) p a] v_{CM}
\label{fors}
\end{equation}
where $E$ is the time average of the energy of the system and $v_{CM}$ is 
the instantaneous velocity of the center of mass defined by
\begin{equation}
v_{CM} = {1 \over N} \sum_{i=1}^N v_i = {1\over N} J_N
\label{vCM}
\end{equation}
where $J_N$ is the number current.

In comparing $J^c_Q$ and $J^l_Q,$ it is useful to first derive an identity
that is valid for the autocorrelation function of any conserved current:
\begin{equation}
\int_0^\infty C(t) dt = \lim_{\tau\rightarrow\infty}\frac{1}{2\tau}
\int_0^\tau\int_0^\tau\langle {\bf J}(t_1)\cdot{\bf J}(t_2)\rangle dt_1 dt_2
= \lim_{\tau\rightarrow\infty}\frac{1}{2}
\bigg\langle\Big[\frac{1}{\sqrt\tau}\int_0^\tau {\bf J}(t) dt\Big]^2\bigg\rangle.
\label{asymp}
\end{equation}
Thus in Eq.(\ref{kubo}), we are interested in the $O(\sqrt\tau)$ part
of $\int_0^\tau {\bf J}_Q(t) dt.$  In particular, in 
Eq.(\ref{gauge})
we have
\begin{equation}
\half p \sum_{i=2}^N (v_{i-1} + v_i)a \equiv (N-1) p a v_{CM}
\end{equation}
similar to Eq.(\ref{fors}), since $\int_0^\tau [v_i(t) - v_{CM}(t)] dt$
is equal to the change in  $x_i(t) - x_{CM}(t),$ which cannot
be $O(\sqrt\tau).$
Note that even though $J_Q(t)$ has a 
power law tail to its autocorrelation function, this tail is cut off for
large time for any finite $N$, as it {\it must\/} be from Eq.(\ref{asymp}) if 
the thermal conductivity is to be finite for that $N$. 

As another useful result, if $\rho$ is the charge density
corresponding to a current ${\bf j}_\rho,$ then
\begin{equation}
\frac{d}{dt}\int {\bf x}\rho({\bf x}, t) d{\bf x} = -\int {\bf x}
\nabla\cdot{\bf j}_\rho({\bf x}, t) d{\bf x} = -\int_S {\bf x} 
{\bf j}_\rho({\bf x}, t)\cdot dS + {\bf J}_\rho(t)
\end{equation}
in which the integral on the right hand side is over the surface of the
system, where it is connected to reservoirs. 

\subsection{Tethered chain}
\label{sec:tethered}
Let the FPU chain of Eq.(\ref{Ham}) be tethered at the ends. This is
accomplished by adding extra fixed particles at the zeroth and $N+1$'th
locations with springs connecting them to their neighbors. With
these boundary conditions, the displacement of the center of mass
in a time interval $\tau\rightarrow\infty$ must be $O(1).$ Then from
Eq.(\ref{asymp}), the energy and heat currents yield the same result in
Eq.(\ref{kubo}), both for the lattice and continuum currents.

From the continuity equation,
\begin{equation}
\frac{d}{dt}\sum (i-1) a e_i(t)  = 
 -\int_0^{(N-1)a} (y -a)\frac{d j^l_E(y,t)}{dy}  dy 
= J^l_E(t) - (N -1) a J_{E;R}(t) 
\label{jelat_vol}
\end{equation}
where $y$ is a continuous coordinate along the lattice, such that
$j^l_E(y)$ is piecewise constant with discontinuities at integer multiples
of the lattice constant $a.$ $J_{E;R}(t)$ is the energy current flowing
into the heat bath at the right end of the system. If we integrate both sides
of this equation over an extremely large time interval $\tau,$ the left hand side does not diverge with $\tau.$ The first term on the right hand side must be 
$O(\sqrt\tau)$ from Eq.(\ref{asymp}), and therefore so must be the second term. 
Keeping only terms that grow as $O(\sqrt\tau),$ we obtain
\begin{equation}
\int_0^\tau J^l_E(t) dt \equiv (N-1) a \int_0^\tau J_{E;R}(t) dt.
\label{jelat_vol1}
\end{equation}
Similarly, since from Eq.(\ref{asymp}) the left hand side of Eq.(\ref{jelat_vol1}) is $O\big(N^{(1+\alpha)/2}\big)$ for large $N$ where $\alpha$ is the heat conductivity
exponent~\cite{llp}, the right hand side must be the same.

We can try to understand the $O\big(\sqrt\tau N^{(1+\alpha)/2}\big)$ scaling of 
the right hand side as follows. On extremely long time scales, $J_{E;R}(t)$ is as likely to be
negative as positive. Thus the right hand side of Eq.(\ref{jelat_vol1})
is $O(\sqrt\tau)$ for large $\tau.$ On the other hand, 
if energy flows into the system from the reservoir to the right, it
increases the local energy density, thereby increasing the likelihood that
energy will flow out shortly afterwords. As $N\rightarrow\infty,$ all the
energy that flows in from the reservoir to the right must eventually flow
out to the same reservoir instead of escaping to the left. Consequently,
$J_{E;R}(t)$ is anticorrelated up to a time scale that diverges with $N.$
This reduces the $O(N)$ dependence one might naively expect for the right
hand side of Eq.(\ref{jelat_vol1}). Although the arguments in this paragraph are qualitative, the $O\big(\sqrt\tau N^{(1+\alpha)/2}\big)$ scaling of the 
right hand side of Eq.(\ref{jelat_vol1}) is derived rigorously in the previous paragraph.

For the continuum energy current, 
\begin{equation}
\frac{d}{dt}\sum e_i(t) x_i = 
\frac{d}{dt}\int_{x_1(t)}^{x_N(t)} e(x,t) x dx = -\int x \frac{d j^c_E(x,t)}{dx}  dx 
= J^c_E(t) +x_1(t) J_{E;L}(t) - x_N(t) J_{E;R}(t) 
\label{jecont_vol}
\end{equation}
where $J_{E;L}$ is the heat current flowing out of the reservoir on
the left hand side of the system. With $x_{1,N}(t) = [0, (N-1)a]
+ \zeta_{1,N}(t)$
\begin{equation}
\int_0^\tau J^c_E(t) dt\equiv (N-1) a \int_0^\tau J_{E;R}(t) dt + 
\int_0^\tau [\zeta_N(t) J_{E;R}(t) - \zeta_1(t) J_{E;L}(t)] dt.
\label{jecont_vol1}
\end{equation}
Compared to Eq.(\ref{jelat_vol1}), the extra term on the right hand
side is also $O(\sqrt\tau)$ for large $\tau.$ However, the missing
factor of $(N-1)$ makes it negligible compared to the first term in the
thermodynamic limit.

One would expect the factors of $\zeta$ in the integrand to destroy
the temporal anticorrelation of $J_{E;L}$ and $J_{E;R}$ so that the last
term should be independent of $N.$ Thus we would expect
\begin{equation}
\frac{\int_0^\infty C_{QQ}^c(t)dt}{\int_0^\infty C_{QQ}^l(t)dt} -1\sim 
N^{-(1+\alpha)}\sim N^{-4/3}.
\label{leading_corr_teth}
\end{equation}
Contrary to this expectation, we will see numerically that the
right hand side is in fact proportional to $1/N,$ as suggested in
Section~\ref{sec:intro}. We must conclude that the correlations between
$\zeta$ and $J_{E;L,R}$ are more subtle than one would naively expect.

\subsection{Constant pressure boundary conditions}
If the FPU chain has a constant pressure $p$ applied to it instead
of being tethered at the ends, the equivalence of the continuum and
lattice heat currents is more delicate. The tethering is removed, and
the pressure effectively adds a term $p(x_N - x_1)$ to the potential
energy in Eq.(\ref{Ham}). The center of mass of the system executes
a random walk due to the fluctuating forces exerted by the heat baths,
moving a distance $O(\sqrt\tau)$ in time $\tau.$ The volume of the system
also fluctuates by $O(\sqrt N).$ However, the lattice description is
essentially the same as for the tethered chain. Since the system at a 
pressure $p$ is equivalent to a system at zero pressure with an extra 
$p(x_i - x_{i-1})$ in the interparticle energy, from Eq.(\ref{gauge})
the continuity equation Eq.(\ref{jelat_vol}) is modified to
\begin{equation}
\frac{d}{dt}\sum (i-1) a [e_i(t) + \half p a (x_i - x_{i-1}) + \half p a (x_{i+1} - x_i)]
= J^l_E(t) - p a \sum_{i=2}^N \half (v_i + v_{i-1}) - (N -1) a J_{E;R}(t)
\label{jelat_pres}
\end{equation}
from which 
\begin{equation}
\int_0^\tau J_Q^l(t) dt \equiv (N-1) a \int_0^\tau J_{E;R}(t) dt
\label{single_bond}
\end{equation}
as before. Note that we could have used $(i - 1 - k) a$ on the left
hand side of Eq.(\ref{jelat_pres}), which would have yielded a linear
combination of $J_{E;R}(t)$ and $J_{E;L}(t)$ on the right hand side
of Eq.(\ref{single_bond}).

Turning to the continuum current, 
\begin{equation}
\frac{d}{dt}\sum e_i(t) x_i = 
J_E^c(t) +x_1 J_{E;L}(t) - x_N J_{E;R}(t) + p x_1 \frac{dx_1}{dt} - p x_N \frac{dx_N}{dt}
\label{jecont_pres}
\end{equation}
where on the right hand side we have included the energy current from
the reservoirs and the work done by the applied pressure. We define
the heat current (integrated over the chain) as 
\begin{equation}
J_Q^c(t) = J_E^c(t) - \Big\{E(t) + p [x_N(t) - x_1(t)]\Big\} v_{CM} 
\label{def}
\end{equation}
where $E(t)$ is the total energy of the system at time $t$ and $v_{CM}$
is the velocity of the center of mass.  This definition is slightly
different from that in Eq.(\ref{fors}); if $E(t)$ and $x_N(t) - x_1(t)$
are replaced with their average values, we recover the earlier expression.
We also use conservation of energy
\begin{equation}
\frac{dE(t)}{dt} = J_{E;L}(t) - J_{E;R}(t) + p \frac{dx_1}{dt} - p\frac{dx_N}{dt}.
\end{equation}
Multiplying this equation by $x_{CM}$, subtracting from
Eq.(\ref{jecont_pres}) and using Eq.(\ref{def}), we have
\begin{eqnarray}
\frac{d}{dt}\sum e_i(t) [x_i - x_{CM}] &=&
J_Q^c(t) + (x_1 - x_{CM}) J_{E;L}(t) - (x_N - x_{CM}) J_{E;R}(t) 
\nonumber\\
&+& 
p (x_1 - x_{CM}) \frac{d}{dt}(x_1 - x_{CM}) - p (x_N - x_{CM}) \frac{d}{dt} (x_N - x_{CM}).
\end{eqnarray}
Therefore
\begin{equation}
\int_0^\tau J_Q^c(t)dt\equiv \half (N-1) a \int_0^\tau[J_{E;R}(t) + J_{E;L}(t)] dt
+ \int_0^\tau (\zeta_N - \zeta_{CM}) J_{E;R}(t) dt + \int_0^\tau (\zeta_{CM} - \zeta_1) J_{E;L}(t) dt.
\label{final_cont}
\end{equation}
As per the discussion after Eq.(\ref{single_bond}), the first term
on the right hand side is equivalent to $\int_0^\tau J_Q^l(t) dt.$
As per the discussion after Eq.(\ref{jelat_vol1}), this is $\sim
\sqrt{\tau N^{1+\alpha}}.$ The last two terms are also proportional
to $\sqrt\tau.$ Unlike for the tethered case, since $\zeta_{1,N} -
\zeta_{CM}$ is $O(\sqrt N),$ it is not clear they can be neglected
even in the thermodynamic limit. Therefore, to compare the lattice
and continuum conductivities with constant pressure boundary conditions,
we turn to numerical simulations in the next section. Note that if we
had used the first definition of the heat current, Eq.(\ref{fors}), 
Eq.(\ref{final_cont}) would have had an extra term on the right hand side
\begin{equation}
\int_0^\tau [E - E(t) - p \zeta_N(t) + p \zeta_1(t)] v_{CM}(t) dt
\end{equation}
with which it is even less clear that the lattice and continuum results 
are equivalent.

\section{Numerical results}
\label{sec:numerics}
In the numerical simulations we set the on-site potential $U_0(x)$ 
in Eq.~(\ref{Ham}) to be zero, and take the interparticle potential to be
\begin{equation}
U(x_{i+1}-x_i) = {1 \over 2}\, (\zeta_{i+1}-\zeta_i)^2 +
{w \over 3} \, (\zeta_{i+1}-\zeta_i)^3 +
{u \over 4}\, (\zeta_{i+1}-\zeta_i)^4 \, ,
\label{Ux}
\end{equation}
where $\zeta_i = x_i - i\,a$ The lattice spacing $a$ is taken to be
unity. We consider $N$ particles of which the first and last are coupled
to heat baths at temperature $T$. Each particle in the interior interacts
with neighbors to the left and to the right, while the first and last
particles only interact with one neighbor. These end particles have
additional forces due to the friction and noise, $-\gamma v_1 + \eta_1$
and $-\gamma v_N + \eta_N,$ and also interact with an extra tethered
`particle' at the zeroth and $N+1$'th sites respectively.  
The noise
satisfies the fluctuation-dissipation theorem $\langle \eta_i(t)
\eta_j(t')\rangle = 2 k_B T \gamma \,\delta(t - t')\, \delta_{ij}$.

We integrated the equations of motion using the second order velocity Verlet
(leapfrog) method for all particles in the interior. For those on the
boundary, which are coupled to the heat baths and so subjected to random noise,
we just used the simple Euler method, so, for example, for particle 1,
\begin{equation}
v_1(t + \delta t) = v(t) + F(\zeta_1 - \zeta_2) \delta t 
- \gamma\, v(t)\, \delta t  + R\, \epsilon(t) ,
\end{equation}
where the force, $F(\zeta_1 - \zeta_2)$, is
equal to $ -d U / d\zeta_1$ with $U$ given by Eq.~(\ref{Ux}),
and $\epsilon$ is a
Gaussian random variable with mean 0 and standard deviation unity. 
The coefficient of the noise, $R$,
is the root mean square fluctuation in the noise integrated
over time $\delta t$, so it is given by
\begin{equation}
R^2 = \int_0^{\delta t} dt \int_0^{\delta t} dt'\ \langle \eta(t) \eta(t')
\rangle = \int_0^{\delta t} dt \int_0^{\delta t} dt'\ 2 k_B T \gamma \,
\delta(t -t')  = 2 k_B T \, \gamma\, \delta t \, .
\end{equation}
We focused on one set of parameters,
$w = 0.5, u = 1.0, \gamma = 0.5, T = k_B = 1.0$, and for the constant pressure
simulations we took $p = 0.1$.

\begin{table}
\caption{
Parameters of the tethered chain simulations for
different values of particle number $N$ and time step $\delta t$. $t_{\rm
equil}$ is the time for equilibration, $t_{\rm meas}$ is the subsequent
time during which measurements are performed, and $n_{run}$ is the
number of ``runs'', where one run comprises the equilibration plus the
measurement time steps.
\label{table:params}
}
\begin{tabular}{|r r| r r r| r r r|}
\hline\hline
 & & \multicolumn{3}{c |}{constant V} 
 & \multicolumn{3}{c |}{constant p} \\
\hline
$N$ & $\delta t$ &
$t_{\rm equil}$ & $t_{\rm meas}$ & $n_{run}$ &
$t_{\rm equil}$ & $t_{\rm meas}$ & $n_{run}$ \\
\hline
16  & 0.10   &   25000 &   50000 & 4000 &   25000 &   50000 & 4000 \\
16  & 0.05   &   25000 &   50000 & 4000 &   25000 &   50000 & 4000 \\
16  & 0.025  &   25000 &   50000 & 4000 &   25000 &   50000 & 4000 \\
16  & 0.0125 &    ---  &    ---  &  --- &   25000 &   50000 & 4000 \\[2mm]
32  & 0.10   &   50000 &  100000 & 4000 &   50000 &  100000 & 4000 \\
32  & 0.05   &   50000 &  100000 & 4000 &   50000 &  100000 & 4000 \\
32  & 0.025  &   50000 &  100000 & 4000 &   50000 &  100000 & 4000 \\
32  & 0.0125 &  100000 &  200000 & 8000 &   50000 &  100000 & 4000 \\[2mm]
64  & 0.10   &  100000 &  200000 & 4000 &  100000 &  200000 & 4000 \\
64  & 0.05   &  100000 &  200000 & 4000 &  100000 &  200000 & 4000 \\
64  & 0.025  &  100000 &  200000 & 4000 &  100000 &  200000 & 4000 \\
64  & 0.0125 &  200000 &  400000 & 8000 &  100000 &  200000 & 4000 \\[2mm]
128 & 0.10   &  200000 &  400000 & 4000 &  200000 &  800000 & 4000 \\
128 & 0.05   &  200000 &  400000 & 4000 &  200000 &  800000 & 4000 \\
128 & 0.025  &  200000 &  400000 & 4000 &  200000 &  800000 & 4000 \\
128 & 0.0125 &  400000 &  800000 & 4000 &  200000 &  800000 & 4000 \\[2mm]
256 & 0.10   &  500000 & 1000000 & 4000 &  500000 & 1000000 & 4000 \\
256 & 0.05   &    ---  &    ---  &  --- & 300000 &  600000 & 8000 \\
256 & 0.025  & 1000000 & 2000000 & 4000 &  500000 & 1000000 & 4000 \\
256 & 0.0125 & 1000000 & 2000000 & 4000 & 1000000 & 2000000 & 4000 \\
\hline
\hline
\end{tabular}
\end{table}

\begin{figure}
\includegraphics[width=9cm]{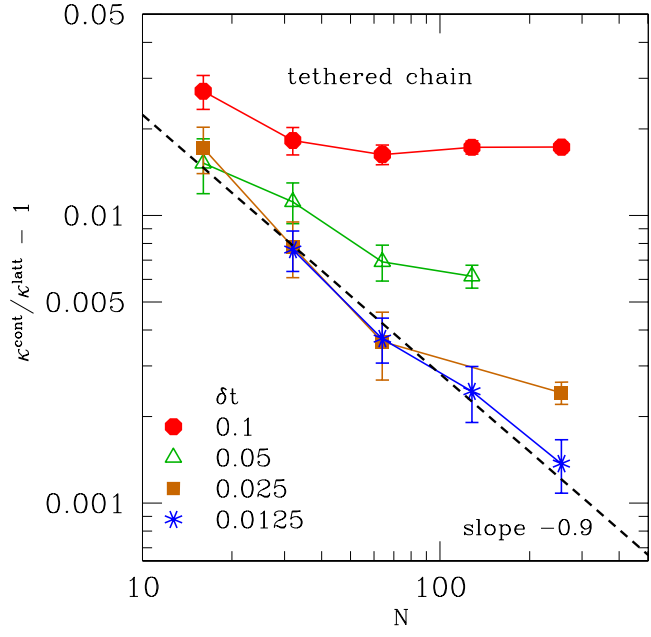}
\caption{
(Color online)
Results for the tethered chain. A log-log plot of 
the ratio of the continuum
thermal conductivity with the heat current defined in Eq.~(\ref{fors})
to the lattice thermal conductivity with the heat current is defined
in Eq.~(\ref{gauge}) minus one.  If the continuum and lattice thermal
conductivities agree in the thermodynamic limit, the results should
tend to zero for $N \to \infty$.  
The best fit has a slope of $-0.9$, close to $-1$ instead of $-4/3,$
which is explained near the end of Section~\ref{sec:intro}.}
\label{fig:constV1}
\end{figure}

\begin{figure}
\includegraphics[width=9cm]{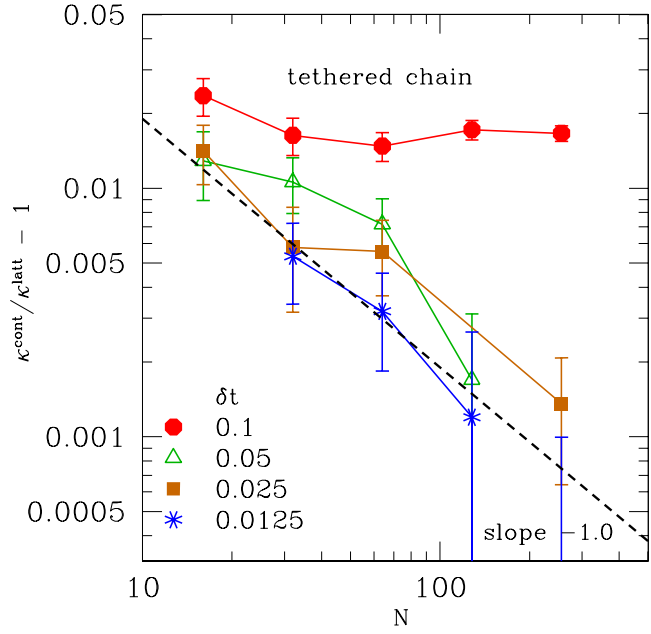}
\caption{
(Color online)
Results for the tethered chain. A log-log plot of 
Same as for Fig.~\ref{fig:constV1}, but using
the continuum heat current given by
Eq.~(\ref{def}).
}
\label{fig:constV2}
\end{figure}

We start the system off with all velocities equal to zero,
and the particles separated by their equilibrium distance $a$, so all the energy
is fed in from the baths at the boundaries. We run for a time $t_{\rm equil}$ to
equilibrate, and then continue for an additional time $t_{\rm meas}$ during
which measurements of the current are made after every time step.
This whole process, both equilibrating and averaging,
is then repeated $n_{\rm run}$ times and the results
averaged. Error bars are estimated from the standard deviation of results from
different runs in the usual way. The values of the parameters used are given in
Table~\ref{table:params}.
The thermal conductivity is obtained from the simulations from
\begin{equation}
\kappa = {1 \over 2\, (N-1)\, T^2\, t_{\rm meas}}
\left\langle \left(\int_0^{t_{\rm meas}} J_Q(t_{\rm equil} + t)
\, d t\right)^2 \right\rangle,
\end{equation}
where $\langle \cdots\rangle$ denotes the average over the
$n_{\rm run}$ runs. 

We investigated the size-dependence of the
difference between the thermal conductivities calculated with the continuum
currents, Eqs.~(\ref{fors}) and (\ref{def}), from that obtained with the
lattice current, Eq.~(\ref{gauge}).
This difference 
is always small, and, especially for the tethered chain simulations,
depends sensitively on the time step, $\delta t$. We shall therefore present
results as a function of both $\delta t$ and $N$.
As discussed at the start of Sec.~\ref{sec:tethered}, 
for a tethered chain in which the heat current
is defined by Eq.~(\ref{fors}),
the terms subtracted from
the energy current are inconsequential for the conductance.

\begin{figure}
\includegraphics[width=9cm]{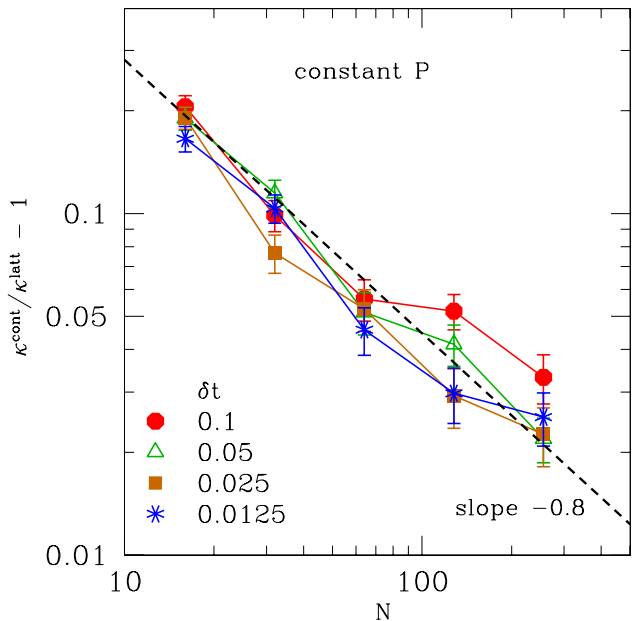}
\caption{
(Color online)
Results for the tethered chain. A log-log plot of 
Results for constant pressure. A log-log plot of the ratio of the continuum
thermal conductivity with the heat current defined in Eq.~(\ref{fors}) to the
lattice thermal conductivity with the heat current defined in
Eq.~(\ref{gauge}) minus one. 
The best fit has a slope of about $-0.8$.
}
\label{fig:constP1}
\end{figure}

\subsection{Tethered Chain}
Figure~\ref{fig:constV1} presents data for the tethered chain
for the ratio of the
thermal conductivity from the continuum current in Eq.~(\ref{fors}) to that
from the lattice current in Eq.~(\ref{gauge}) minus 1, as a function of system
size $N$, for different values of the time step $\delta t$.  If the continuum
and lattice thermal conductivities agree in the thermodynamic limit, the
results should tend to zero for $N \to \infty$.  It is clearly essential to
extrapolate the results to small values of $\delta t$, and when one does so,
the result is close to the expected $1/N$ dependence. The best fit of the data
for the smallest value of $\delta t$ gives $1/N^\eta$ with $\eta \simeq 0.9$,
but the small difference in the value of the exponent $\eta$ from one is
probably due to corrections to scaling which are not completely negligible for
this range of sizes.

\begin{figure}
\includegraphics[width=9cm]{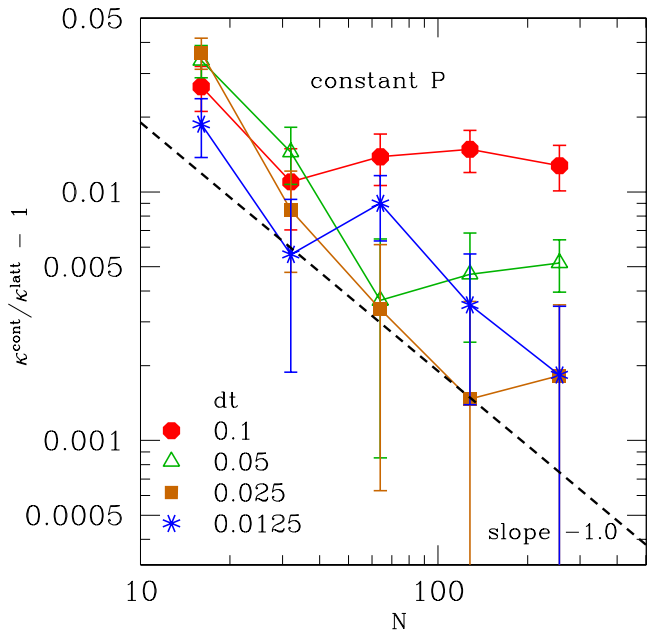}
\caption{
(Color online)
Results for the tethered chain. A log-log plot of 
Same as Fig.~\ref{fig:constP1} but using the continuum heat current
given in Eq.~(\ref{def}). The error bars are sufficiently large that it is not
possible to estimate the size dependence reliably.
}
\label{fig:constP2}
\end{figure}

In Fig.~\ref{fig:constV2} we show data similar to Fig.~\ref{fig:constV1}, but
using the second definition of the continuum current, in Eq.~(\ref{def}). The
difference in thermal conductivities is smaller, so it is harder to get good
statistics on it, but it also seems to decay with size like $1/N$.
This conclusion disagrees
with the naive expectation of Eq.~(\ref{leading_corr_teth}),
but is consistent with the physical
argument given in the introduction. 

\subsection{Constant Pressure Chain}
For the constant pressure chain,
Fig.~\ref{fig:constP1} plots 
the ratio of
the thermal conductivity using the continuum heat 
current in Eq.~(\ref{fors}) to that using the lattice current in
Eq.~(\ref{gauge}) minus 1. For reasons that are not clear to us, the
dependence on time step is much smaller than for the corresponding
tethered chain results
in Fig.~\ref{fig:constV1}. A fit to the data gives a size dependence of
$N^{-\eta}$ with $\eta \simeq 0.8$. Taken naively, this is a faster decay than
the 
$1/N^{1/2}$ dependence expected for constant pressure, as discussed in the
introduction. However, the error bars on the exponent are sufficiently large that any $1 > \eta > \half$ is possible. It is also possible that larger sizes are needed to see
the asymptotic size dependence.  We note that the relative 
uncertainty in the length of the system is of order $1/N^{1/2}$ and it seems
surprising to us that difference in thermal conductivities could be smaller
than this asymptotically. Despite these uncertainties,
the relative difference in thermal conductivities is small,
even for constant pressure,
and appears to vanish in the thermodynamic limit.

Finally, Fig.~\ref{fig:constP2} is similar to Fig.~\ref{fig:constP1} but uses
the alternative definition of the head current in Eq.~(\ref{def}). There is
still a non-zero difference compared with the lattice conductivity
but it is smaller, so the error bars are larger, as a result of which
it is not possible to
reliably estimate any functional form for the difference. 

\section{Conclusion}
In this paper, we have shown that the continuum and lattice versions of
the heat current yield different results for the heat conductivity of a
finite oscillator chain with Langevin baths
through the Green-Kubo formula. The results using
the continuum and lattice currents are different for a finite chain.
Since the thermal conductivity obtained using the lattice current is
exact~\cite{kdn}, this implies that results using the continuum current
are approximate even though this current
is apparently exact.  Numerically, if the error is fit to a form $\sim
1/N^\eta$ for a chain of $N$ particles, we obtain $\eta \lesssim 1$ for a
chain that is tethered just beyond its end points.  We can argue that the
error is at least partly because the length of the chain fluctuates, and
therefore must be bounded below by $\sim 1/N;$ our numerical results are
close to and slightly above this lower bound. The same argument
would yield a lower bound to the error of $\sim 1/\sqrt N$ for a chain at
constant pressure. Surprisingly, though,
the numerical results for such a chain show an error
that decays {\it faster}, almost as $1/N,$ although the asymptotic large
$N$ exponent may approach 0.5. A small change in the definition of
the continuum heat current, which is benign in the thermodynamic limit,
reduces the error substantially.
%
%
%


\begin{thebibliography}{}
\bibitem{foot:abh} 
It is easy to prove that the time average of the two currents is the 
same when a small temperature difference is applied across the system.
Here we are considering the conductivity as obtained from the equilibrium
correlations. As discussed later in the paper, this involves calculating 
the $\sim O(\sqrt\tau)$ fluctuations in the current integrated over a time
$\tau,$ which is more delicate than the $O(\tau)$ part of the out 
of equilibrium integral.
\bibitem{foot1} The tethering is applied one lattice site beyond the ends
of the chain, so that the length of the chain fluctuates by $O(1).$
\bibitem{foot:hydro} In a hydrodynamic description, the displacement of
the particles from their lattice positions is a function of the density,
which is a conserved quantity in the longitudinal interpretation; in the
transverse interpretation it is a broken symmetry (even though there is
no long range order in the chain). Therefore in both interpretations,
the displacement is associated with a hydrodynamic variable, though the
cause is different.
\end{thebibliography}
\end{document}